\documentclass[preprint,showpacs,showkeys,preprintnumbers,amsmath,amssymb]{revtex4}

\usepackage{graphicx}
\usepackage{dcolumn}
\usepackage{bm}

\begin{document}

\preprint{APS/TES}

\title{Electron-Phonon Coupling on the NbSi Transition Edge Sensors}

\author{Shu-chen Liu}
 \email{Shu-Chen.Liu@csnsm.in2p3.fr}
\author{Stefanos Marnieros, Louis Dumoulin, Youri Dolgorouki, Laurent Berge, Sophie Collin}%
\affiliation{%
Lab CSNSM, Bat108 Universit${\acute{e}}$ de Paris-Sud XI, Orsay 91405, France
}%


\date{\today}

\begin{abstract}
We have built an electron-phonon coupling model to describe the behavior of the ${Nb_xSi_{1-x}}$ transition edge sensor (TES) bolometers, fabricated by electron-beam coevaporation and photolithography techniques on a 2-inch silicon wafer. The resistance versus temperature curves of several sensors with different thickness are measured with different bias currents, ranging from 200 nA to 10 ${\mu}$ A, and the electron-phonon coupling coefficient and the electron-phonon thermal conductance are calculated herein. Our values are quite comparable with those in metallic TES samples of other groups using different measurement methods, while we are using the transition region of our TES sample to calculate the electron-phonon coupling interaction. 
\end{abstract}

\maketitle

\section{\label{sec:level1}Introduction}

We have developed bolometer matrices with transition edge sensors (TES) in NbSi. This novel design increases the sensitivity for each pixel, reduces the phonon noise below ${10^{-17}W/\sqrt{Hz}}$ \cite{fabrication} and allows multiplexed readout by SQUIDs \cite{multiplexer,book:SQUID}. The sensitive bolometer matrices and the transition edge sensors (TES) could be applied to measure the temperature anisotropies of the cosmic microwave background (CMB) radiation, which was created about 400,000 years after the Big Bang when the universe became transparent. The last-scattered photons of CMB radiation by free electrons become polarized by Thompson scattering. By study of the CMB temperature and polarization anisotropies, the initial condition of our Universe can be revealed and cosmological theories such as inflation or gravitational waves could be tested.

Our prototype 23 pixel ${Nb_xSi_{1-x}}$ bolometer matrices (with x=0.14) are fabricated on a 2-inch silicon wafers using electron-beam co-evaporation and photolithography techniques. The superconducting transition temperature can be adjusted by the Nb concentration “x” and the thickness of the ${Nb_xSi_{1-x}}$ thin film. Several sensors of the NbSi TES matrices are measured and observed to undergo metal to superconductor transition at 75 mK and 110 mK for the thickness of 20 nm and 50 nm, respectively. The resistance versus temperature curves, slightly shift to lower temperature values when a higher excitation current is applied.

We use an ``electron-phonon coupling'' model to describe our TES\textquoteright s behavior. The electron-phonon coupling coefficient, ${g_{e-ph}}$, is calculated from the slope of the fitted curve of the electrically dissipated power into the TES versus (${T_e^5-T_{ph}^5}$), where ${T_e}$ is the electron temperature and ${T_{ph}}$ is the phonon temperature of the NbSi.  We compare our data results with other types of thermometers.

\section{\label{sec:level1}Sample Preparation}

\subsection{\label{sec:level2}Basic Structure of One Pixel of Bolometer Matrices }

The basic structure of one pixel of bolometer matrices comprises a radiation absorber, a thermometer and a weak thermal link to the cold bath. To absorb radiation we can use feed horns \cite{horn,horn_clover}, antennas \cite{antenna,antenna_microstrip} or direct absorption by thin films (bismuth or copper) \cite{Bi,MoAuTES}. Incident power of photons (P) is measured by a thermometer, such as superconducting transition edge sensors (TES, e.g. NbSi \cite{LT25Stef} , Mo/Au \cite{MoAuTES}, Ti, etc.) or high impedance Anderson insulators (e.g. NbSi \cite{StefanosPRL}, NTD Ge \cite{NTDGe}, Si:P, etc.). The most commonly used sensors in bolometers are superconducting transition edge sensors and high impedance Anderson insulators. TES have usually higher sensitivities and are well adapted to SQUID readout electronics, while Anderson insulators use traditional JFETs. The characteristic of the transition edge sensors is the sharp superconducting transition in the resistance versus temperature curves, while for Anderson insulators the resistance versus temperature curves have exponential rising and higher impedance at lower temperatures. Each pixel of the bolometer matrices is thermally isolated from a cold bath and since the temperature change equals to photons incident power over thermal conductance (${\Delta T=P/G}$), the thermal conductance must be small to optimize sensitivity. Silicon nitride membrane is used to thermally isolate sample holder weakly coupled to a heat sink in the bolometer matrices, because silicon nitride material has very poor thermal conductance. 

Our developed bolometer matrices in NbSi without silicon nitride membranes are used for three purposes. First, the niobium silicon thin film is used for absorption of radiation with or without antennas. Second, for the creation of thermal decoupling and third, as a temperature sensor. Direct radiation absorption needs impedance matching of NbSi to the 377 ohm vacuum impedance. This is possible by adjusting the NbSi thickness and composition. Thermal decoupling is based on the electron-phonon decoupling into the NbSi sensor so the membranes are not necessary for thermal insulation. Either TES NbSi or Anderson Insulator NbSi can be used to scan the temperature. 

\begin{figure*}
\includegraphics{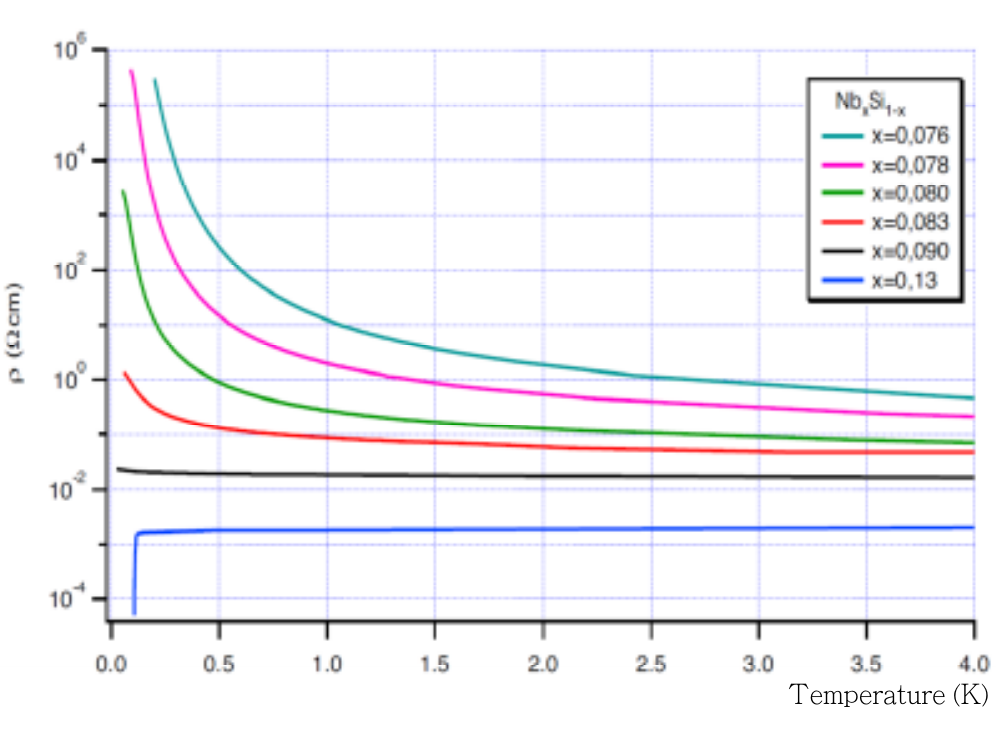}
\caption{\label{NbSi} Resistivity versus temperature curves of ${Nb_xSi_{1-x}}$ thin film show two phase transitions: from superconducting state, to metallic state and finally to insulating state, as the composition of Nb, x, decreases.}
\end{figure*}

Niobium silicon can be either TES or Anderson insulator, depending on the concentration of the niobium and silicon and the sample thickness. For three-dimensional amorphous ${Nb_xSi_{1-x}}$ thin film (e.g. 100 nm), if x ${<9\%}$, it would demonstrate Anderson Insulator state; if ${9\% < x < 12\%}$, it would demonstrate metallic state; if ${x > 12\%}$, it would demonstrate superconducting state. When we decrease the composition x, the niobium silicon thin film shows two phase transitions, from superconducting state, to metallic state, and finally to insulating state (Fig. \ref{NbSi}).

\begin{figure*}
\includegraphics[width=6.375in,height=3in]{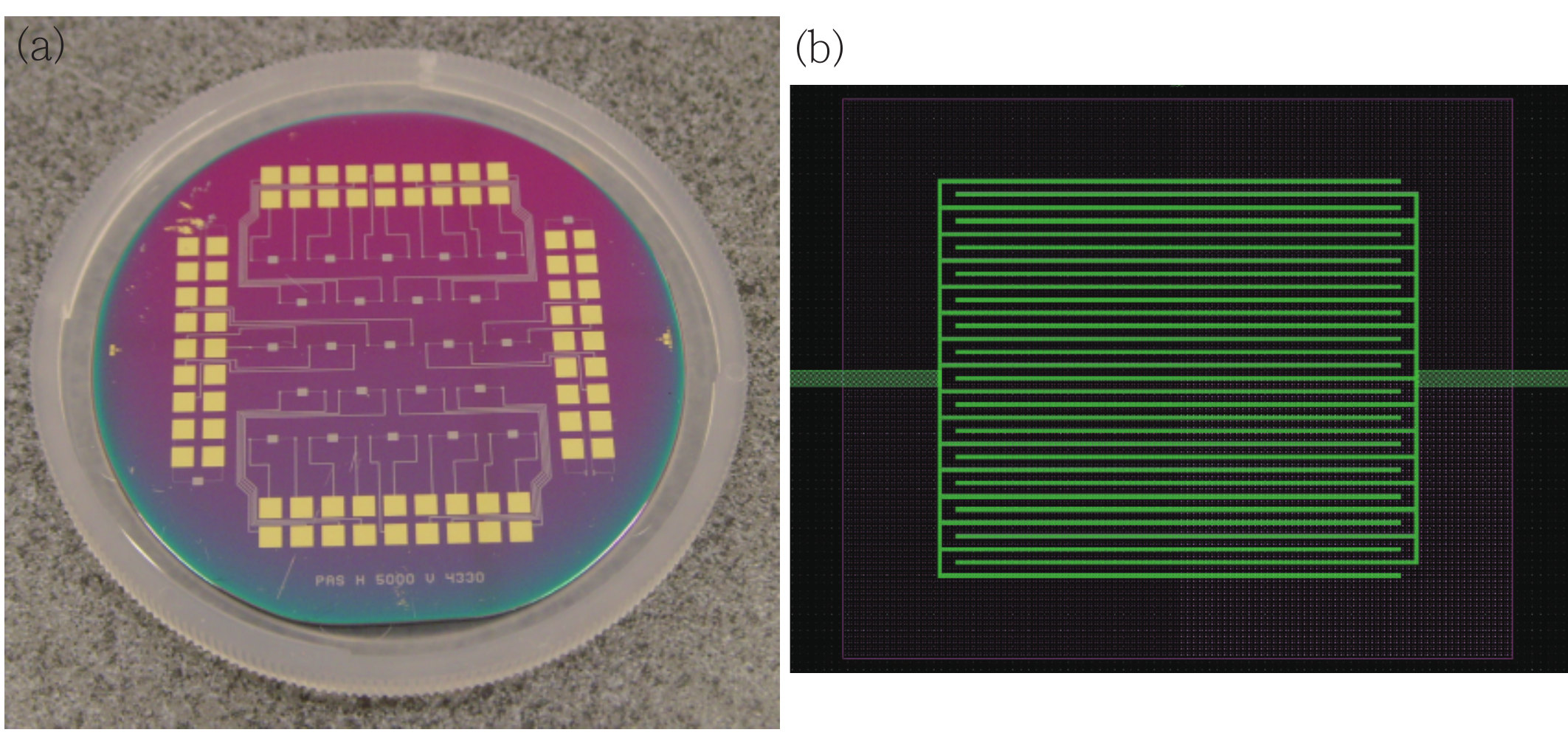}
\caption{\label{bolometer} (a) 23 pixels of NbSi bolometer matrices. (b) A pixel from layout of mask with thermometers, electrodes and grids. In the center region is the interlacing Nb electrodes (green color) on NbSi thin film (pink color).}
\end{figure*}

Our prototype of 23 pixel matrices of superconducting NbSi alloy transition edge sensors (Fig. \ref{bolometer} (a)) is composed of ${Nb_xSi_{1-x}}$ (niobium silicon) thin film with adjustable thickness and composition (x), niobium leads and niobium electrodes interlacing on NbSi thin film (Fig. \ref{bolometer} (b)) with adjustable numbers of electrodes and distances, and gold contact pad. The size of of each pixel, NbSi thin film covered in the range of Nb electrodes, is ${300 \mu m \times 600 \mu m}$.

The superconducting transition temperatures, ${T_c}$, and the normal resistances can be fine-tuned by adjusting the Nb concentration, x, and the thickness, d. The number of the Nb electrodes and the interleaving distances are delicately designed in order to achieve a normal state NbSi resistance appropriate for dc-SQUID readout. We compare results from two ${Nb_{0.14}Si_{0.86}}$ TES samples with thickness of 20 nm and 50 nm. 

\subsection{\label{sec:level2}Fabrication Process of NbSi Bolometer Matrices}

The procedure steps of our simplified, but highly reliable microfabrication are described as in the followings \cite{fabrication}:

\begin{enumerate}

\item 1 ${\mu m}$ thick SiN deposition by PECVD (Plasma Enhanced Chemical Vapor Deposition) on a two-inch silicon wafer. This layer is for electrical insulation.

\item ${Nb_xSi_{1-x}}$ co-evaporation (x=0.14, 20-50 nm): NbSi thin film is manufactured by electron-beam co-evaporation of Nb and Si.

\item Photolithography to form ${Nb_xSi_{1-x}}$ bolometer matrices.

\item Nb evaporation (50 nm) and photolithography to form Nb tracks and electrodes.

\item Au evaporation and photolithography: a gold layer (100-150 nm) is deposited on the wafer to form the square electrical contact pads.

\end{enumerate}

\section{\label{sec:level1}Measurements and Results}

\subsection{\label{sec:level2}Experimental Setup and Methods}

\begin{figure}
\includegraphics[width=6in,height=3.33in]{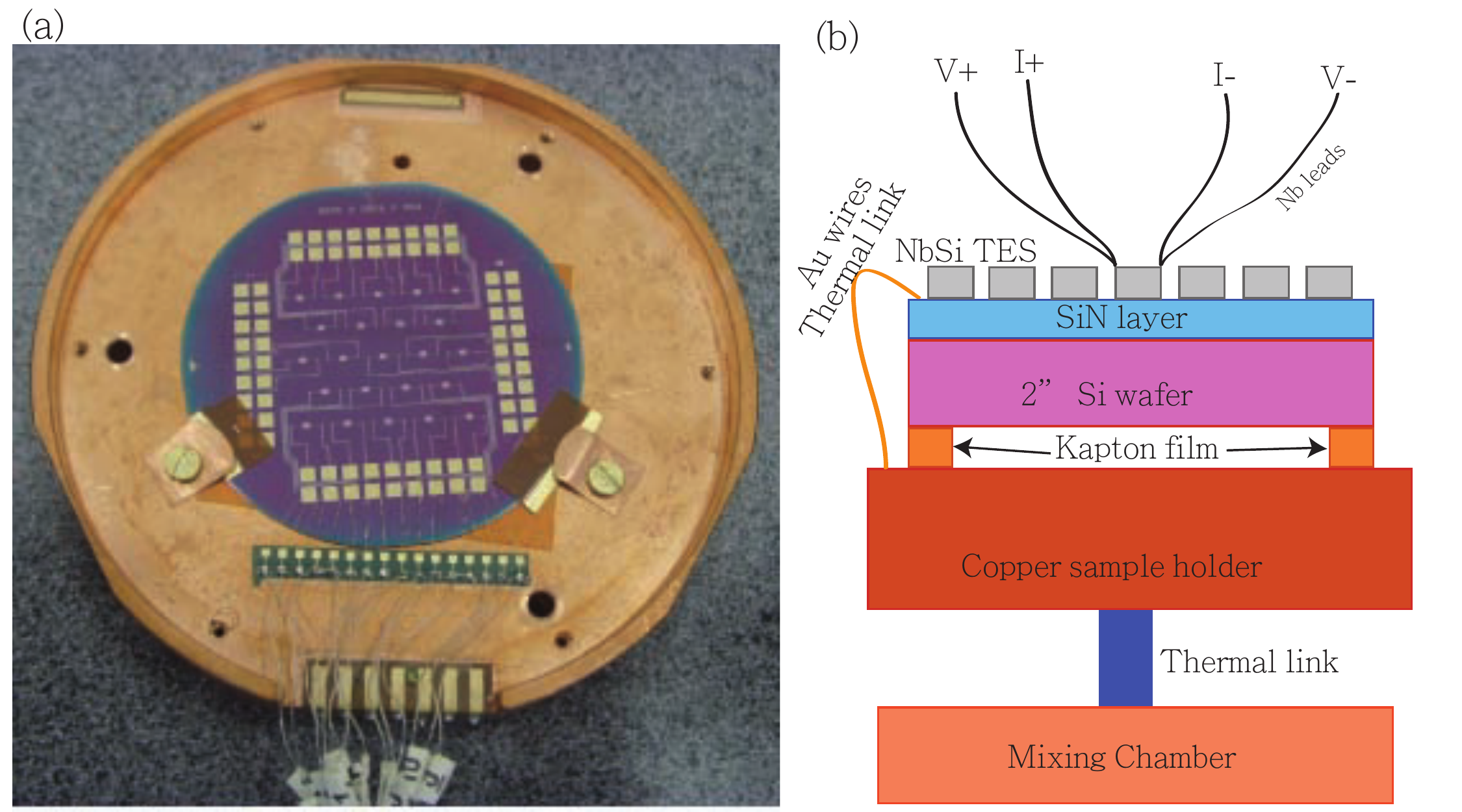}
\caption{\label{Model} (a) Mounting and wiring the Si wafer with 23 pixels of TES's on the copper sample holder. (b) Schematic of the sample setup.}
\end{figure}

In order to eliminate the leads resistances, we use the ``four point measurement'' method to measure the resistance versus temperature curves of ${Nb_{0.14}Si_{0.86}}$ TES thin film with different excitation bias currents ranging from 200 nA to 10 ${\mu}$ A. Fig. \ref{Model} (a) shows the 2-inch wafer, with 23 pixels of NbSi TES sensors fabricated on it, is mounted on a pure copper sample holder with a layer of Kapton film to thermally/electrically insulate between the wafer and the copper disk. Aluminum thin wires of 25 ${\mu}$ m are ultrasonically bonded to electrically connect the gold contact pads between different pixels in series. The aluminum thin wires would become superconducting with zero resistance and are poor thermal conducting at very low temperatures. 25 ${\mu}$ m thin gold wires are ultrasonically bonded to electrically and thermally connect the gold contact pads between the pixels on the wafer and the circuit board on the sample holder which connects the external electronics readout systems. Some extra gold wires create a direct thermal link between the silicon substrate and the copper sample holder. The gold wires have very good thermal conductance and help to cool down the silicon wafer along with the pixels to very low temperatures. The wafer is affixed to the sample holder by two screws and two pieces of copper sheets, which are thermally/electrically isolated from the wafer by small Kapton sheets.

Fig. \ref{Model} (b) shows the schematic setup of the sample. Four-point measurement is used to acquire the R(T) curves of the NbSi TES. The NbSi TES pixels to be characterized are bonded with four gold wires for constant bias current supply and for the voltage drop measurement. SiN membrane has very poor thermal conductance to thermally and electrically insulate the pixels from the silicon wafer. Under the wafer is the Kapton film and then the Copper disk sample holder which thermally links to the mixing chamber of the dilution refrigerator. However, a direct thermal link between the silicon substrate and the copper sample holder is created by the extra gold wires.

\subsection{\label{sec:level2}An Electron-Phonon Coupling Model}

\begin{figure}
\includegraphics{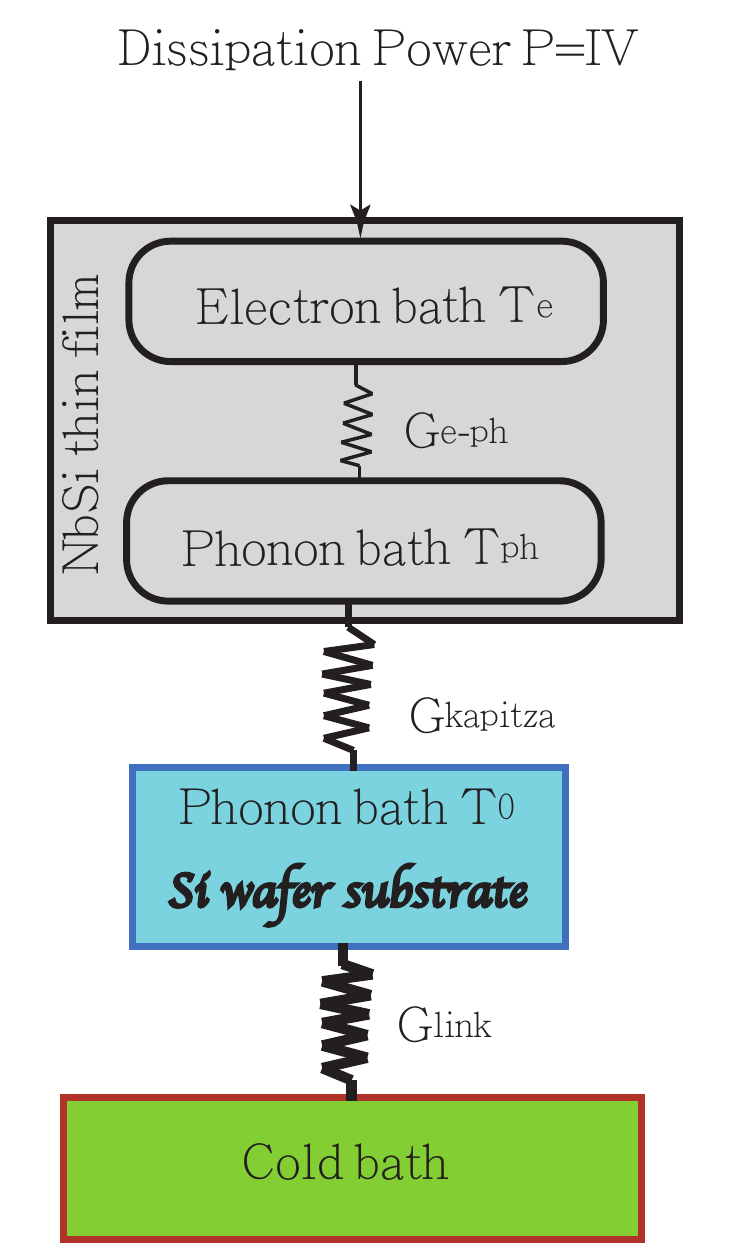}
\caption{\label{e-phModel} Schematic of the electron-phonon coupling model.}
\end{figure}

We apply an ``electron-phonon coupling'' model to describe our TES\textquoteright s behavior (Fig. \ref{e-phModel}). The lowest bias current is applied to the NbSi TES, for example 200 nA, the power is less than 16 femto-Watt; we acquire the sensor resistance (R) versus temperature (T) curves. At this time, the electron and phonon baths and the substrate are in thermal equilibrium and the temperatures are equal (${T_e \approx T_{ph} \approx T_0}$). Then higher dc bias currents are applied to the NbSi TES, increasing the temperature of the NbSi electron bath by Joule heating effect, ${P=I^2R}$. Then the energy is transferred to the NbSi phonon bath via the electron-phonon interaction (${G_{e-ph}}$) and to the Si substrate by the Kapitza interface thermal conductance, ${G_{Kapitza}}$. The silicon wafer is thermally connected to the cryostat cold bath by the extra gold wires which has the thermal conductance, ${G_{link}}$. 

We have measured ${G_{Kapitza}\approx 15 nW/K}$ while ${G_{e-ph}\approx 2.5 nW/K}$. We assume that the thermal conductance between the silicon wafer substrate and the cold bath is much higher than the Kapitza thermal conductance, much higher than the electron-phonon thermal conductance (${G_{link}>>G_{Kapitza}>>G_{e-ph}}$); therefore ${T_e}$ is greater than ${T_{ph}}$, which is almost equal to ${T_0}$ (${T_e>T_{ph}\approx T_0}$). 

\begin{figure*}
\includegraphics{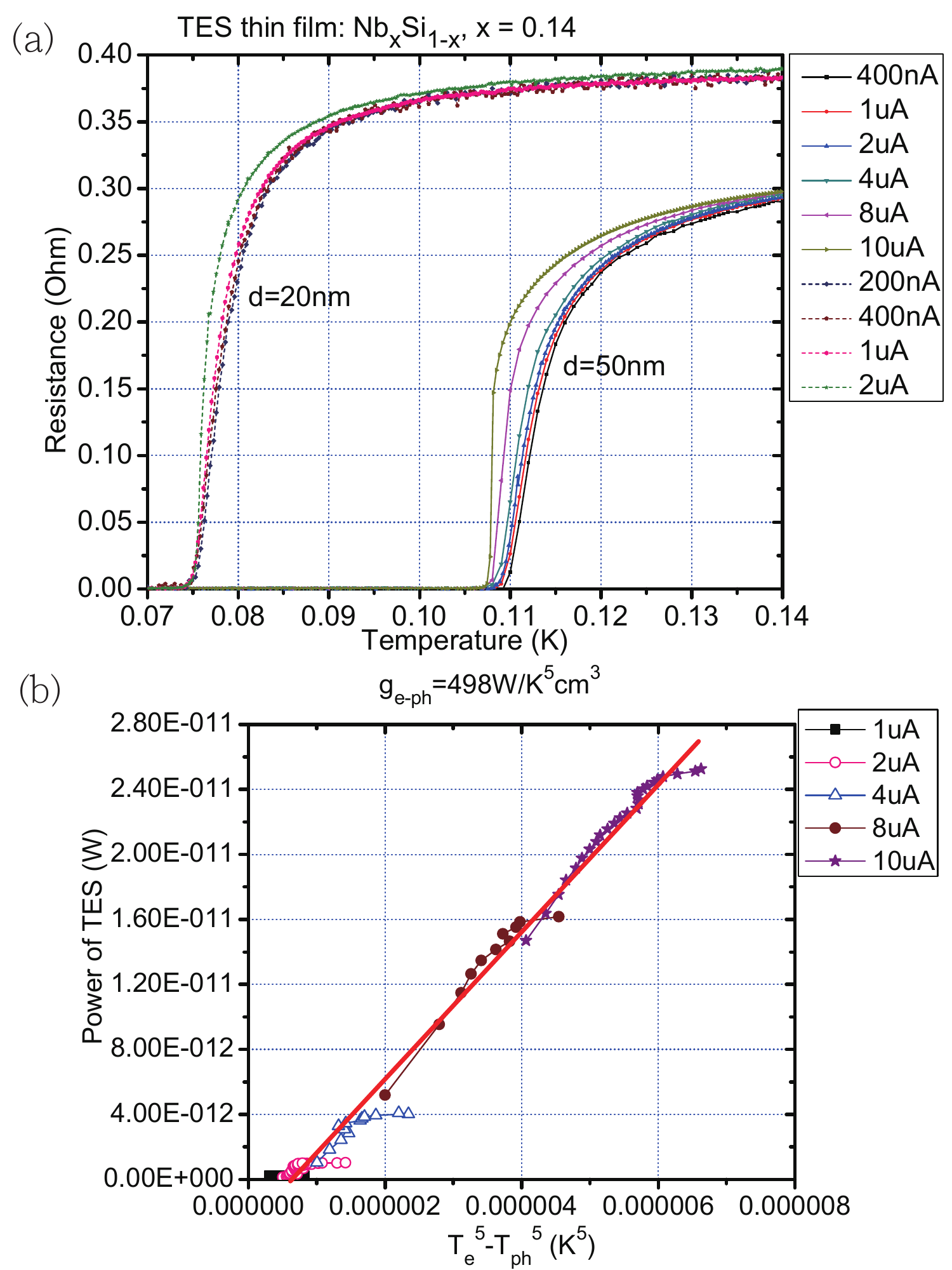}
\caption{\label{R(T)_geph} (a) Resistance versus temperature curves of ${Nb_{0.14}Si_{0.86}}$ TES for different excitation currents (thickness of 20 nm and 50 nm). (b) The electron-phonon coupling coefficient (${g_{e-ph}}$) is given by the slope of the dissipated electrical power versus (${T_e^5-T_{ph}^5}$). The data with ${R<0.1R_n}$ and ${R>0.8R_n}$ are removed (${R_n}$ is the normal state resistance of the TES).}
\end{figure*}

Fig. ~\ref{R(T)_geph} (a) shows the resistance versus temperature curves of ${Nb_{0.14}Si_{0.86}}$ TES thin film with thickness 20 nm and 50 nm, respectively. We observe that it undergoes metal to superconductor transitions at the transition temperature 75 mK and 110 mK for 20 nm and 50 nm thick NbSi TES thin film, respectively (and metal to insulator transition for lower concentration of Nb, e.g. x=0.08). As higher excitation bias currents are applied, the resistance shifts to lower and lower temperature values (Fig. ~\ref{R(T)_geph} (a)).

On the NbSi thin film, the electrical power is dissipated to the electron bath of the bolometer thin film. The electron bath temperature is ${T_e}$. The heat is transfer to the phonon bath with temperature ${T_{ph}}$ via the electron-phonon thermal conductance, ${G_{e-ph}}$. The relation between the electron temperature, ${T_e}$, and the phonon temperature, ${T_{ph}}$, for a given electrical power, P, is described by this formula \cite{StefanosPRL,NTDGe} :

\begin{equation}
\frac{P}{\Omega}=g_{e-ph}(T_e^{\beta}-T_{ph}^{\beta}),
\end{equation} where ${\Omega}$ is the active volume of NbSi TES thin film, ${g_{e-ph}}$ is the electron-phonon coupling constant, ${\beta}$=5 in the case of metals. 

The thermal conductance between the electrons and the phonons is given by the derivative of the power with respect to the electron bath temperature:
\begin{equation}
G_{e-ph}=(\partial P/\partial T_e)_{T_{ph}}=g_{e-ph}\beta T_e^{(\beta-1)}
\end{equation}

\begin{figure*}
\includegraphics{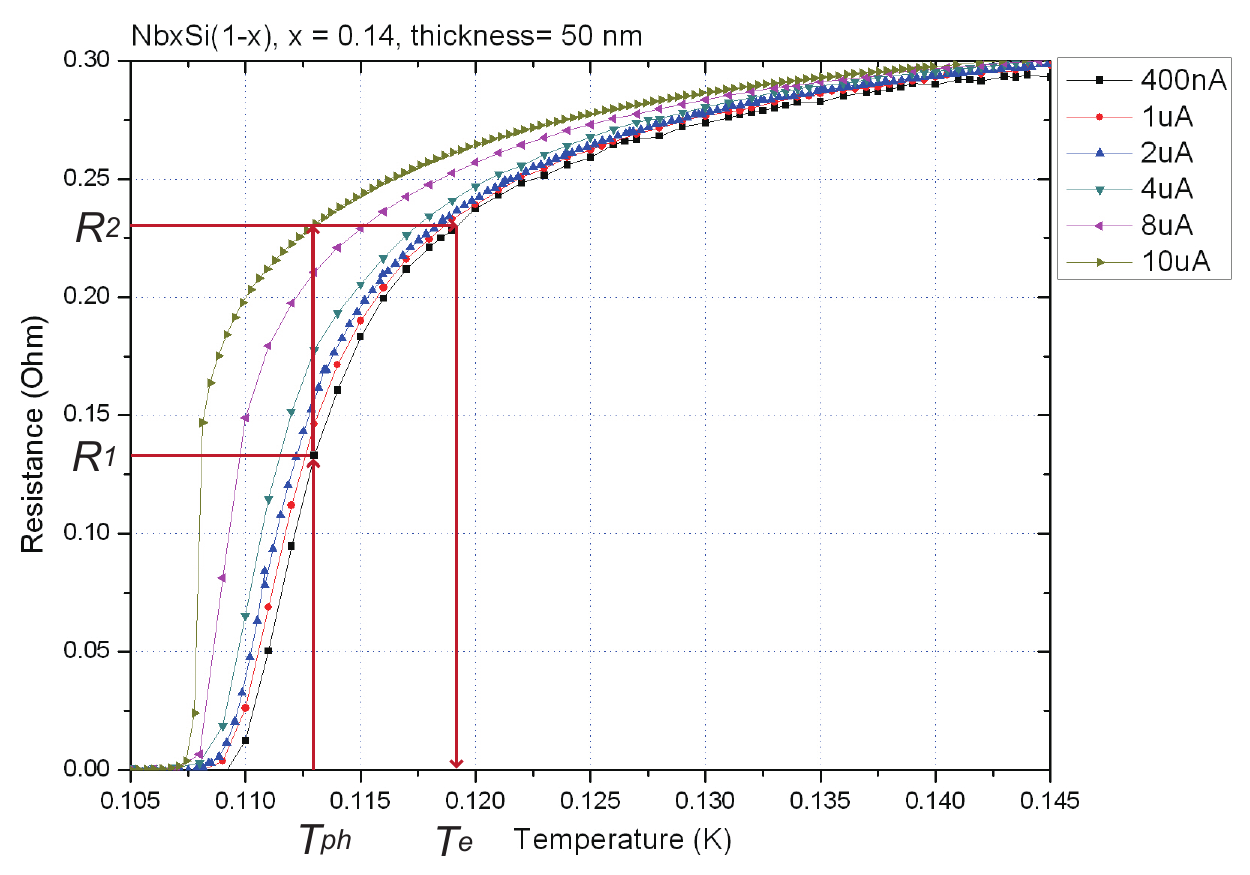}
\caption{\label{Te_Tph} ${T_e}$ is interpolated from the R(T) curve (=R(${T_e}$)) as the lowest bias current is applied, while ${T_{ph}}$ is interpolated from the R(T) curves (=R(${T_{ph}}$)) as the higher bias currents are applied.}
\end{figure*}

The data of Fig. ~\ref{R(T)_geph} (a) are acquired by fixing the bias current and changing the phonon temperature which is controlled by a heater on the mixing chamber and monitored by a calibrated standard thermometer attached to the copper sample holder. To extract ${T_e}$, we suppose the resistance of our TES depends only on the temperature of the electron bath (R=R(${T_e}$)). Under high bias, ${T_e}$ increases due to the Joule heating of the electrons and R(${T_e}$) is rising up, up to the normal state resistance (${R_n}$). As we can see from Fig. ~\ref{Te_Tph}, at the temperature ${T_{ph}}$ the resistance of the NbSi TES is ${R_1}$ as the lowest bias current, 400 nA, is applied. When the higher bias current is applied, say 10 ${\mu}$A, the dissipated electrical power (${P=I^2R=10R_1}$nW) heats up the electron bath of the NbSi TES thin film. Consequently, the resistance increases up to ${R_2}$, which corresponds to the temperature ${T_e}$ on the R(T) curve at the lowest bias current. This was based on the fact that the electron bath temperature ${T_e}$ at the resistance ${R_2}$ is equivalent to the temperature of the phonon and electron baths in thermal equilibrium in the NbSi thin film with the resistance ${R_2}$ at the lowest bias current when there is negligible heat dissipation on the same NbSi TES sensor.
The electron-phonon coupling coefficient, ${g_{e-ph}}$, is calculated from the slope of the fitted curve of the electrically dissipated power into the NbSi TES sample versus (${T_e^5-T_{ph}^5}$) (see Fig. ~\ref{R(T)_geph} (b)). Here the electrically dissipated power (P) into the NbSi TES is extracted from ${P=I^2R}$, where ${I}$ is the higher bias currents, and ${R}$ is the corresponding resistance extracted from 0.1 ${R_n}$ to 0.8 ${R_n}$ (${R_n}$: normal state resistance). ${T_e}$ is the electron bath temperature in the NbSi thin film sample and is interpolated from the R(T) curve (=R(${T_e}$)) as the lowest bias current is applied. ${T_{ph}}$ is the phonon bath temperature in the NbSi thin film sample and is  interpolated from the R(T) curves (=R(${T_{ph}}$)) as the higher bias currents are applied (Fig. ~\ref{Te_Tph}). 

We measured ${g_{e-ph} \approx 498W/K^5cm^3}$, and ${G_{e-ph} \approx 3.28 \times 10^{-9} W/K}$ for the thickness of 50 nm NbSi thin film (Fig. ~\ref{R(T)_geph} (b)); for the thickness of 20 nm thin film, ${g_{e-ph} \approx 558W/K^5cm^3}$, and ${G_{e-ph} \approx 3.18 \times 10^{-10} W/K}$. When the thickness of the NbSi thin film is increased from 20 nm to 50 nm (by 2.5 times), ${g_{e-ph}}$ is almost constant in agreement with the electron-phonon coupling model.

\section{\label{sec:level1}Discussion}

\subsection{\label{sec:level2}Comparison}

\begin{table*}
\caption{\label{tab:table1} Comparison of the electron-phonon coupling constant or thermal conductance with different materials from other groups.}
\begin{ruledtabular}
\begin{tabular}{cccccc}
 Category & Materials & Size & ${g_{e-ph}}$ & ${G_{e-ph}}$ & ${\beta}$ \\
 & & & (${W/K^{\beta}cm^3}$) & (${W/K}$) & \\ \hline
 TES & ${Nb_{0.14}Si_{0.86}}$ \footnotemark[1] & ${300\mu m \times 600 \mu m \times 50 nm}$ & 498 & ${3.28 \times 10^{-9}}$ & 5 \\
 & & ${300\mu m \times 600 \mu m \times 20 nm}$ & 558 & ${3.18 \times 10^{-10}}$ & 5 \\
 Anderson insulator & ${Nb_{0.083}Si_{0.917}}$\footnotemark[2] & ${100\mu m \times 100 \mu m \times 100 nm}$ & 100 & $5 \times 10^{-11}$\footnotemark[3] & 5 \\
 Anderson insulator & NTD Ge \footnotemark[4] & ${1 mm \times 1 mm \times 0.2 mm}$ & 40 & & 6 \\
 & & ${1 mm \times 1 mm \times 0.2 mm}$ & 6.5 & & 5.5 \\
 TES & Ti/Au \footnotemark[5] & ${1 mm \times 1 mm \times 20 nm}$ & 3000 & & 4 \\
 TES & Ir \footnotemark[6] & ${75 \times 75  \mu m^2}$ & & ${1.2 \times 10 ^{-11}}$ & \\
 TES & Mo/Au \footnotemark[7] & ${700 \mu m \times 0.35 \mu m}$ & & ${19 \times 10 ^{-15}}$ & \\
 & & ${700 \mu m \times 0.5 \mu m}$ & & ${72 \times 10 ^{-15}}$ & \\
 TES & Ti/Au \footnotemark[8] & ${150 \mu m \times 150 \mu m \times 75 nm}$ & 2000 & ${2 \times 10 ^{-9}}$ & 5 \\
\end{tabular}
\end{ruledtabular}
\footnotetext[1]{S.C. Liu and S. Marnieros, et al.}
\footnotetext[2]{S. Marnieros, et al.\cite{fabrication}} 
\footnotetext[3]{At 100 mK} 
\footnotetext[4]{N. Wang, et al.\cite{NTDGe}}
\footnotetext[5]{R. Horn, et al.\cite{TiTES}}
\footnotetext[6]{D. Bagliani, et al.\cite{IrTES}}
\footnotetext[7]{M. Kenyon, et al.\cite{MoAu_IR}}
\footnotetext[8]{P. Korte (SRON)}
\end{table*}

For comparison, the electron-phonon thermal coupling constant for a 100 nm thick, 100${\mu m \times 100 \mu m}$ ${Nb_xSi_{1-x}}$ (x=0.083) Anderson insulator thin film is ${g_{e-ph}=100 W/K^5cm^3}$, or ${G_{e-ph}=5 \times 10^{-11}}$ W/K (${G_{e-ph}}$ is proportional to the film volume and ${T^4}$) at 100 mK \cite{LT25Stef}. At 20 mK, ${G_{e-ph}=1 \times 10^{-13}}$ W/K, which drops to 0.002 times of that value at 100 mK. In this thermal decoupling model, the energy is deposited into the electron bath directly, and there exists natural thermal decoupling between electrons and phonons at low temperatures. Another example for an ``electron-phonon decoupling'' model applied on the 100 nm thick ${Nb_xSi_{1-x}}$ (0.075 ${<}$ x ${<}$ 0.09) Anderson insulator thin films, gives ${g_{e-ph}=(130\pm 30) W/K^5cm^3}$ which is constant and independent of temperatures \cite{StefanosPRL}. In these two cases, electron-phonon decoupling effects near the MIT (Mott-Anderson metal-insulator transition) is electron-electron interaction assisted variable range hopping transport without phonons at low temperatures. 

For the neutron-transmutation-doped Ge (NTD Ge) thermometers with the dimension ${1mm \times 1mm \times 0.2mm}$ \cite{NTDGe}, the temperature independent electron-phonon thermal coupling constant is measured to be ${g_{e-ph} = 0.0080 W/K^6}$ (${\beta =6}$) or ${g_{e-ph} = 40 W/K^6cm^3}$ for one thermistor with direct link to a heat sink, and ${g_{e-ph} = 0.0013 W/K^{5.5}}$ (${\beta =5.5}$) or ${g_{e-ph} = 6.5 W/K^{5.5}cm^3}$ for the other thermistor without the thermal sink. A model includes both variable-range-hopping conduction and hot-electron effects (or called electron-phonon decoupling thermal model) can describe the performance of the temperature dependence of zero bias resistance, the I-V curves and the dynamic behavior of the NTD Ge thermistors very well.
 
For the voltage-biased gold-titanium (Au-Ti) bi-layer transition edge sensors (1mm$\times$1mm) on silicon wafers, the electron-phonon coupling was measured to be ${\kappa /VT^3=3\times 10^9 W/m^3K^4=3\times 10^3W/cm^3K^4}$ as ${R/R_n\to 1}$, where ${\kappa}$ is the thermal conductance from the bolometer electrons to the Si phonons and ${V}$ and ${T}$ are the volume and transition temperature of the bolometer \cite{TiTES}. The transition temperature (${T_c}$=200-800 mK) of the Au-Ti bolometers is a function of the Au/Ti thickness ratio and Ti films were typically 20 nm thick. 

For the 75${\times}$ 75 ${\mu m^2}$ Ir transition edge sensors (TES) detector as a detector \cite{IrTES} to measure one single photon of 450 nm wavelength at the temperature of 100-120 mK, the electron-phonon thermal conductance is measured to be ${G_{e-ph}=1.2 \times 10^{-11}W/K}$.

For the Mo/Au bilayer thin film TES \cite{MoAu_IR}, the thermal conductance G is measured to be 72fW/K to 19 fW/K, depending on the SiN support beam structures. The thermal conductance is a function of ${T^{1/2}}$ showing the effective elastic scattering of the acoustic phonon modes.    
In the group of SRON, the Ti/Au bilayer TES thin film with the size of ${150\mu m \times 150 \mu m \times 75 nm}$ is measured to have the electron-phonon coupling coefficient ${g_{e-ph}=2000W/K^5cm^3}$, and the electron-phonon thermal conductance ${G_{e-ph}}$ is measured to be ${2 \times 10^{-9}}$ W/K.

Our values are comparable to those of other groups using different measurement methods, as concluded in Tables~\ref{tab:table1}. In our method, we are using the transition region of our TES sample to calculate the electron-phonon coupling interaction. In general we observe that TES has higher electron-phonon coupling constant or thermal conductance than Anderson insulators. 

\subsection{\label{sec:level2}Paraconductivity}

\begin{figure}
\includegraphics[width=5.5in,height=3.75in]{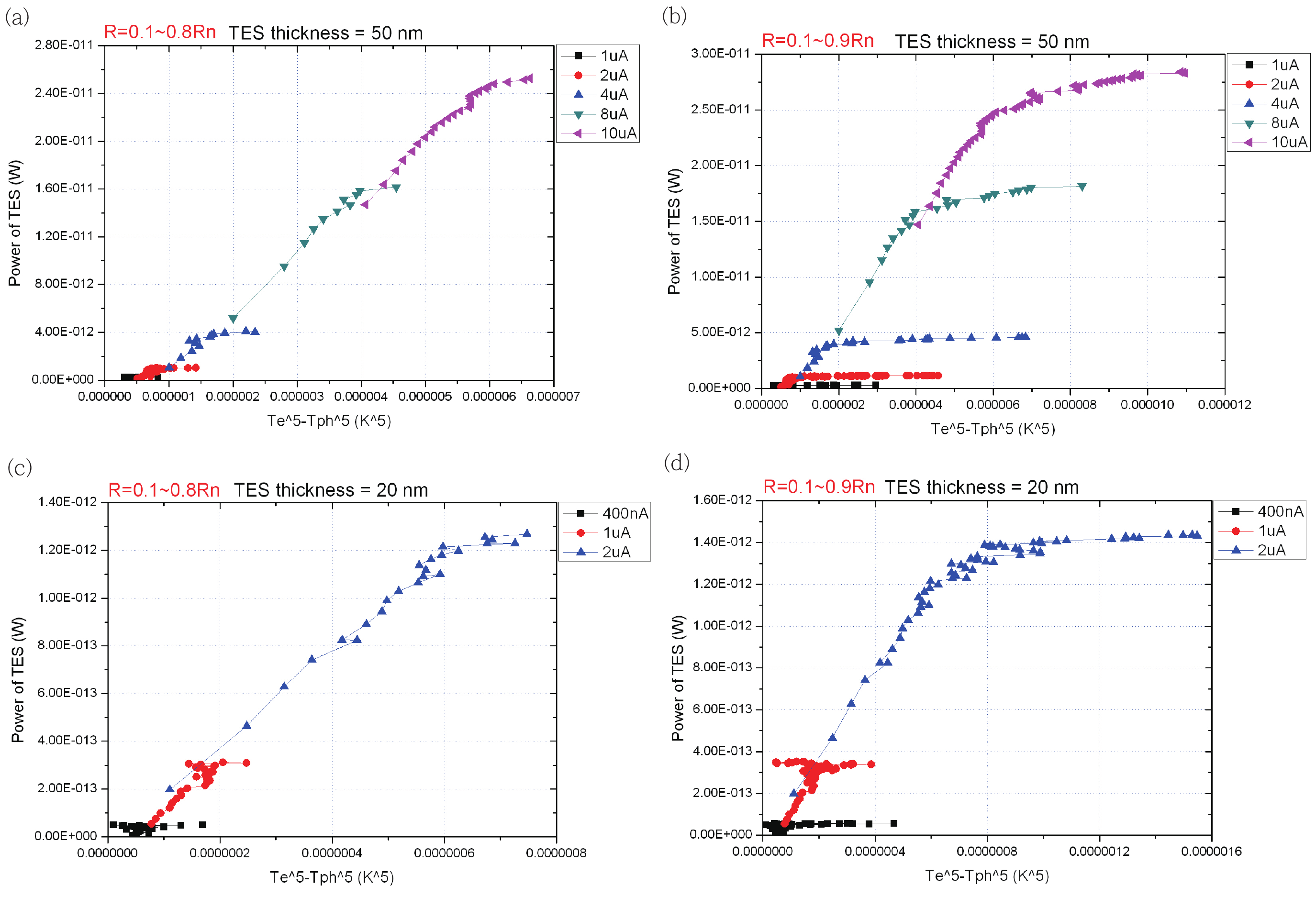}
\caption{\label{paraconductivity} Electrically dissipated power, ${P=I^2R}$, versus the differences of the electron and phonon bath temperatures in the NbSi thin film samples by the exponent order of 5, ${T_e^5-T_{ph}^5}$: (a) R=0.1-0.8 ${R_n}$, ${Nb_{0.14}Si_{0.86}}$ TES thin film thickness 50 nm; (b) R=0.1-0.9 ${R_n}$, ${Nb_{0.14}Si_{0.86}}$ TES thin film thickness 50 nm; (c) R=0.1-0.8 ${R_n}$, ${Nb_{0.14}Si_{0.86}}$ TES thin film thickness 20 nm; (d) R=0.1-0.9 ${R_n}$, ${Nb_{0.14}Si_{0.86}}$ TES thin film thickness 20 nm.}
\end{figure}

In calculating the electron-phonon coupling coefficients from the slopes of the electrically dissipated power, ${P=I^2R}$, versus the 
differences of the electron and phonon bath temperatures in the thin film samples by the exponent order of 5, ${T_e^5-T_{ph}^5}$, we have extracted the data points of the resistance (R) versus temperature (T) curves from 0.1 ${R_n}$ to 0.8 ${R_n}$ along with the corresponding electron or phonon temperatures (see Fig. \ref{paraconductivity} (a) and (c)). If we extract the data points from 0.1 ${R_n}$ to 0.9 ${R_n}$ instead, from the resistance (R) versus temperature (T) curves along with the corresponding temperatures, we have the power versus (${T_e^5-T_{ph}^5}$) curves like Fig. \ref{paraconductivity} (b) and (d). By comparing Fig. \ref{paraconductivity} (a) and (b), (c) and (d), it is very clear to see that between ${R=0.8R_n}$ and ${R=0.9R_n}$, the slopes of the power versus (${T_e^5-T_{ph}^5}$) curves change to be much smaller and almost flat out. In this region, for almost about the same electrical power dissipated into the TES sample, the difference between the electron bath temperature and the phonon bath temperature in the thin film by the exponent order of 5, could range, for example, from ${7.5\times10^{-7}}$ to ${16\times10^{-7} K^5}$, and change the magnitude as large as by twice or three times. The physical meanings of this phenomenon in this region with a depletion of the normal state resistance slightly above the transition temperatures (so-called paraconductivity) could be the thermal fluctuations due to the presence of short-lived thermally activated Cooper pairs. Paraconductivity, also called excess conductivity is defined as the difference between the conductivity at the transition region and the normal state conductivity, ${\sigma'=\sigma-\sigma_0}$.

\begin{figure}
\includegraphics[width=5.5in,height=4in]{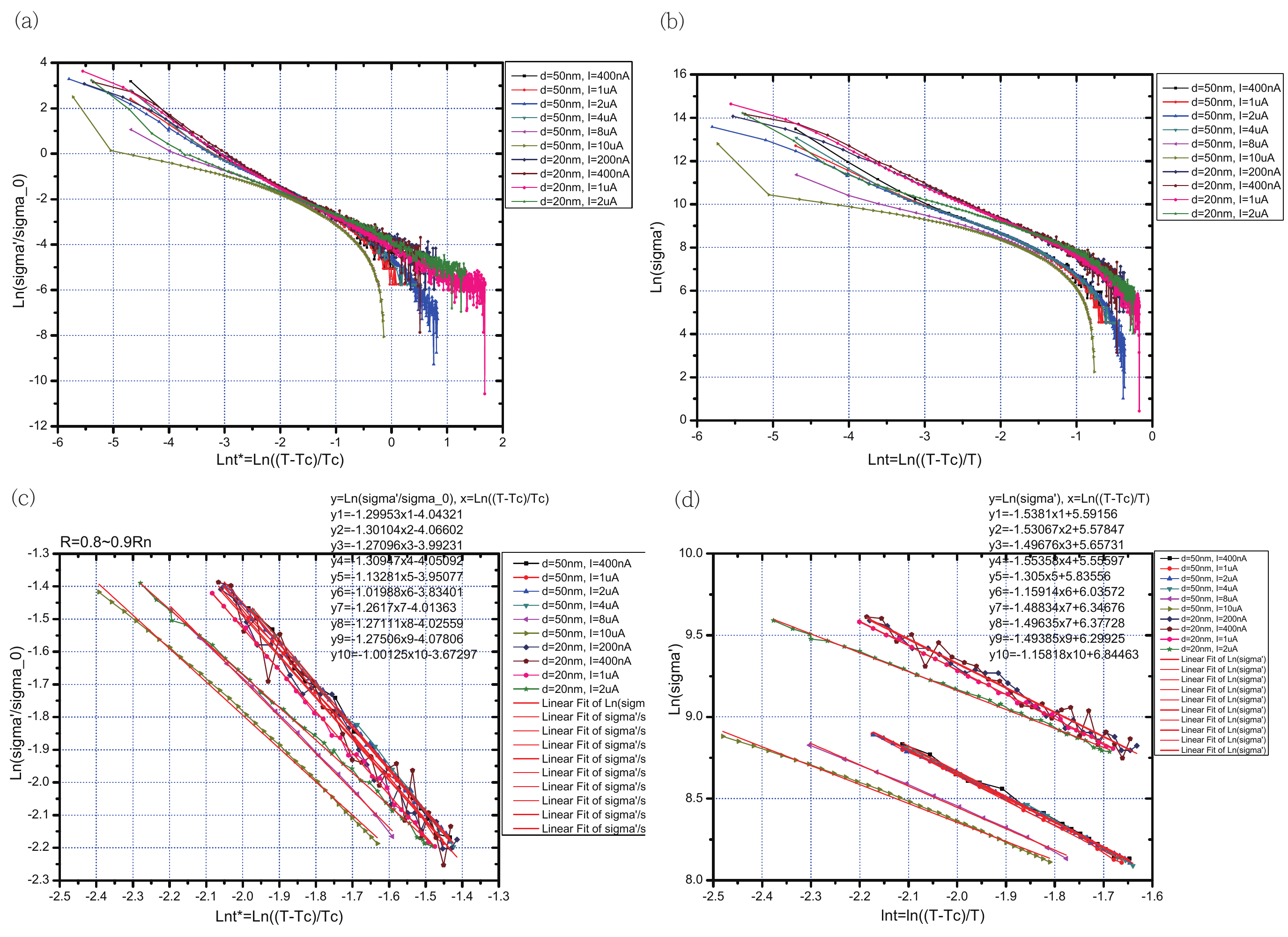}
\caption{\label{LNparacond} (a) ${Ln(\sigma'/\sigma_0)}$ versus ${Ln[(T-T_c)/T_c]}$ for the whole range of R. (b) ${Ln(\sigma')}$ versus ${Ln[(T-T_c)/T]}$ for the whole range of R. (c) ${Ln(\sigma'/\sigma_0)}$ versus ${Ln[(T-T_c)/T_c]}$ for ${R=0.8 \sim 0.9R_n}$. (d) ${Ln(\sigma')}$ versus ${Ln[(T-T_c)/T]}$ for ${R=0.8 \sim 0.9R_n}$.}
\end{figure}

We plot the natural logarithm of the normalized paraconductivity (${\sigma'/\sigma_0}$) versus the natural logarithm of ${t^*=(T-T_c)/T_c}$, i.e. ${Ln(\sigma'/\sigma_0)}$ versus ${Ln[(T-T_c)/T_c]}$ for the whole range of R (see Fig. \ref{LNparacond} (a)), and the natural logarithm of the paraconductivity (${\sigma'}$) versus the natural logarithm of ${t=(T-T_c)/T}$, i.e. ${Ln(\sigma')}$ versus ${Ln[(T-T_c)/T]}$ for the whole range of R (see Fig. \ref{LNparacond} (b)). We found from both graphs that the slopes and curvatures behave differently for different bias currents and thickness, and also vary at different x values. If we only plot for the region of ${R=0.8 \sim 0.9R_n}$ (see Fig. \ref{LNparacond} (c) and (d)), we found the linear relationships for all various bias currents and different thickness, and the higher the bias current, the less negative the slope would become. In other words, in Fig. \ref{LNparacond} (c):

\begin{eqnarray}
Ln(\frac{\sigma'}{\sigma_0})=A\times Ln(\frac{T-T_c}{T_c})+B,\\
\frac{\sigma'}{\sigma_0}=e^B \times [\frac{T-T_c}{T_c}]^A,
\end{eqnarray} where A is the slope and B is the intercept. As the bias current rises up, the slope, A, changes from ${\sim}$ -1.3 to -1.0 for both 20 nm and 50 nm thick NbSi TES thin films. Similarly, in Fig. \ref{LNparacond} (d):

\begin{eqnarray}
Ln(\sigma')=A'\times Ln(\frac{T-T_c}{T})+B',\\
\sigma'=e^{B'} \times [\frac{T-T_c}{T}]^{A'},
\end{eqnarray} where A' is the slope and B' is the intercept. As the bias current rises up, the slope, A', changes from ${\sim}$ -1.5 to -1.16 for both 20 nm and 50 nm thick NbSi TES thin films. Since the temperature dependence of ${\sigma'}$ is found to be ${(T-T_c)^{-(4-d)/2}}$, where d (=1,2,3) is the dimensionality of the system \cite{Tinkham}, we reveal from our data that the dimension increases from about one dimension to two dimension as the bias current is increased, which is obviously violate the reality situation.

Theoretical prediction has given the electrical field dependence of the paraconductivity that the Cooper pairs get accelerated along the fluctuation's size by applied electrical field and increase their kinetic energy up to suppress the fluctuation itself \cite{paraconductivity2}. The extra kinetic energy that the Cooper pairs have earned departs them from the stable thermal equilibrium with thermal phonons. Therefore, the electron bath temperature arising from the unstable pairing and pair-breaking Cooper paris with high thermal fluctuations at the paraconductivity zone has much farther way to reach the thermal equilibrium with the phonon bath temperature. The discussion about the temperature and magnetic-field dependence of the paraconductivity of three-dimensional amorphous superconductors has been reported \cite{paraconductivity1,book:fluctuation}. The physics mechanisms existing in the region of ${R=0.8 \sim 0.9R_n}$ for the reasons of the deviation from the power conductance law (${P\propto (T_e^5-T_{ph}^5)}$), and whether it is related with the paraconductivity and how, are still under further investigation.

\subsection{\label{sec:level2}Sensitivity}

The sensitivity is defined as:
\begin{equation}
\alpha =dR/dT
\end{equation}
(in ${\Omega /K}$), where R is the resistance of the thermometer sensor, and T is the temperature. Another definition of the dimensionless sensitivity is:

\begin{equation}
\alpha '=\frac{dlogR}{dlogT}=(\frac {T}{R})(\frac {dR}{dT}).
\end{equation}

\begin{figure}
\includegraphics[width=5.85in,height=4.13in]{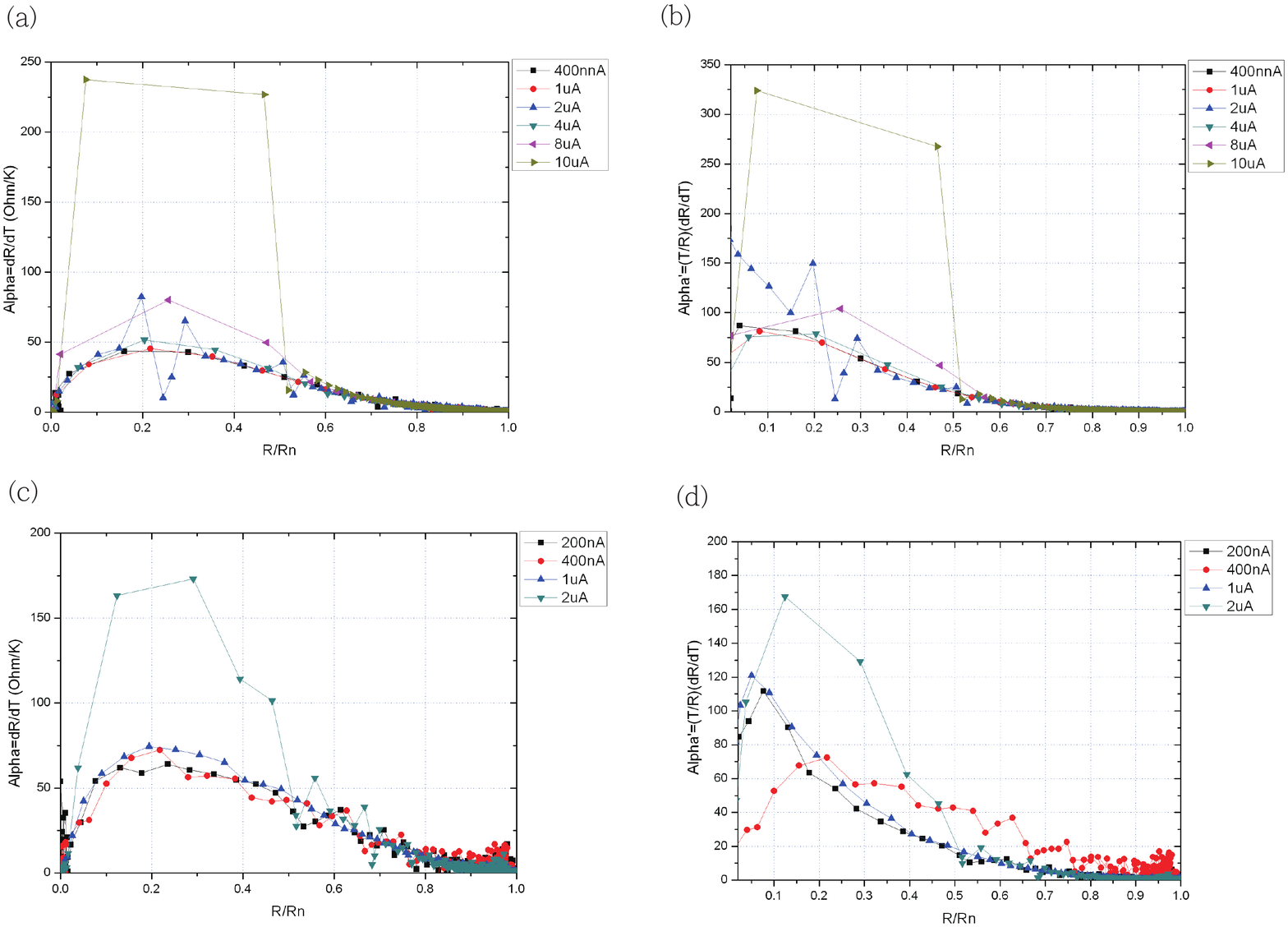}
\caption{\label{Sensitivity} (a) and (c): Sensitivity (${\alpha = dR/dT}$) of the NbSi TES sensor with thickness of 50 nm and 20 nm, respectively, versus the normalized resistance (${R/R_n}$); (b) and (d): dimensionless sensitivity (${\alpha '=(T/R)(dR/dT)}$) of the NbSi TES sensor with thickness of 50 nm and 20 nm, respectively, versus the normalized resistance (${R/R_n}$).}
\end{figure}

The sensitivities, ${\alpha}$ and ${\alpha'}$, are calculated via the above equations and from the measurement of the resistance versus temperature curves of 20 nm and 50 nm thick ${Nb_{0.14}Si_{0.86}}$ TES thin film with different excitation bias currents (Fig. \ref{R(T)_geph} (a)). In Fig. \ref{Sensitivity} (a) and (c), the sensitivity (${\alpha = dR/dT}$) shows the maximum values 50--240 and 70--175 at the normalized resistance (${R/R_n}$) around 0.2 ${\sim}$ 0.3 for 50 nm and 20 nm thickness of NbSi TES sensors, respectively. The higher the excitation bias current, the more sensitive the sensor becomes. NbSi TES sensor of 20 nm thickness has consistently higher sensitivity than the one of 50 nm thickness does. Fig. \ref{Sensitivity} (b) and (d) shows the dimensionless sensitivity (${\alpha '=(T/R)(dR/dT)}$) of the NbSi TES sensor with 50nm and 20 nm thickness, respectively, versus the normalized resistance (${R/R_n}$). ${\alpha '}$ demonstrates higher sensitivity as ${R/R_n \to 0}$ and decays to zero as ${R/R_n \to}$ 1. For the thickness 50 nm, the maximum sensitivity is ranging from 75 to 325 for bias current from 400 nA to 10 ${\mu}$A. For the thickness 20 nm, the maximum sensitivity is ranging from 70 to 170 for bias current from 200 nA to 2 ${\mu}$A. It is very comparable with other groups. In general, the higher the bias current, the higher the sensitivity. NbSi TES sensor of 20 nm thickness has higher dimensionless sensitivity than the one of 50 nm thickness; the higher the excitation bias current, the higher the ${\alpha '}$ values.

For Ti/Au-bilayer TES with the Bi/Cu absorber, maximum ${\alpha'}$ is 130 [Korte's group, SRON]. For Mo/Au bilayer TES, maximum ${\alpha'}$ is less than 40 for an Au absorber and 100 with Au/Bi absorber fabricated on a ${Si_3N_4}$ membrane \cite{MoAuTES}. In comparison, we have maximum ${\alpha'=75 \sim 325}$ and ${70 \sim 170}$, depending on the excitation bias currents, for NbSi TES sensor of 50 nm and 20 nm thickness, respectively. Therefore, we conclude that our fabricated NbSi TES sensors have comparable or even better performance in terms of the sensitivity measurement.

\subsection{\label{sec:level2}Dc-SQUID Circuitry and Noise Spectra}

\begin{figure*}
\includegraphics[width=5.85in,height=4.13in]{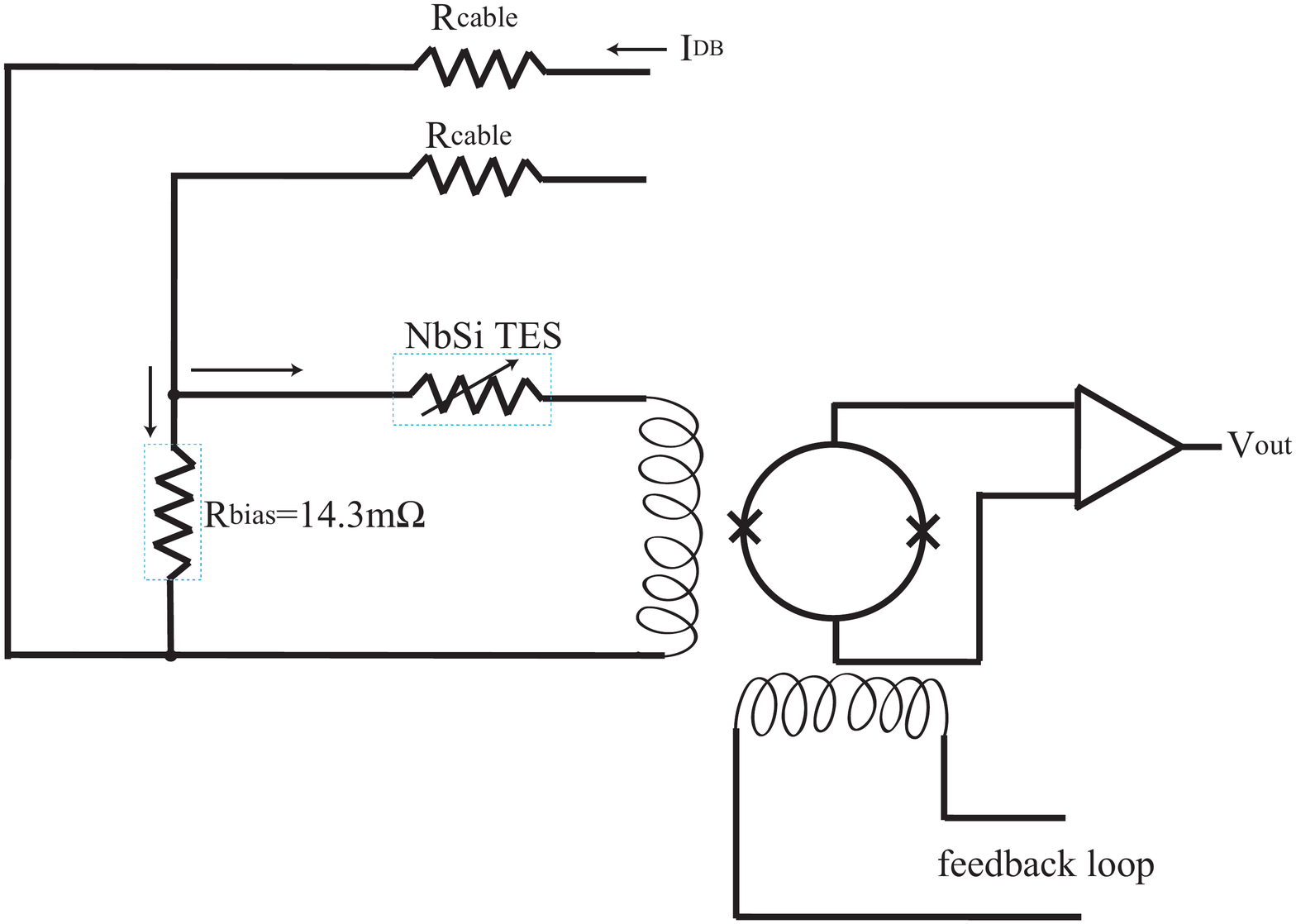}
\caption{\label{SQUID} Schematic circuitry for the dc-SQUID to readout the TES sensor.}
\end{figure*}

The schematic circuitry of the dc-SQUID (Supracon, Jessy) for the measurement of the noise spectra and the I-V characteristics of the NbSi TES sensors is shown in Fig. \ref{SQUID}. The dc-SQUID is equipped in contact with the 1 K pot of the dilution refrigerator, while the NbSi TES is settled on the copper sample holder with thermal link to the mixing chamber. At low enough temperatures, the detector bias current, ${I_{DB}}$ is supply to the circuit and split to flow into both NbSi TES sensor (resistance ${R_{TES}}$) and the bias resistor (resistance ${R_{bias}}$). Note: ${R_{TES}}$ (for the normal resistance above ${T_c}$)${>>R_{bias}}$. Therefore, the current flowing through the NbSi TES is:
\begin{equation}\label{I_tes}
I_{TES}=I_{DB} \times \frac{R_{bias}}{R_{bias}+R_{TES}},
\end{equation}
and the current flowing through the bias resistor is:
\begin{equation}
I_{bias}=I_{DB} \times \frac{R_{TES}}{R_{bias}+R_{TES}}.
\end{equation}
When the NbSi TES sensor undergoes the superconducting transition as the temperature is lowered to below ${T_c}$, the resistance of the sensor (${R_{TES}}$) would change (from ${\approx 400 m \Omega}$ for 20nm thick NbSi TES and from ${\approx 300 m \Omega}$ for 50nm thick NbSi TES to zero), as well as the current flowing through it (from ${I_{TES}=0.0345I_{DB}}$ for 20nm thick NbSi TES and ${0.0455I_{DB}}$ for 50nm thick NbSi TES to 1${I_{DB}}$). The inductance of the inductor (impedance${<<R_{TES}}$ for the normal resistance above ${T_c}$) connecting to the NbSi TES in series would change as the current flowing thought it changes, so would the magnetic flux inside the inductor. The induced current flowing through the SQUID, acting as a flux-voltage transducer, would give the voltage readout after amplified by the amplifier (gain =6000), and is proportional to the external flux (or the current flowing through the inductor coil) after the electronics integrator. The feedback loop is coupled to the SQUID for the compensation of the external flux change in the SQUID to conserve the total flux and to bring back the SQUID to the optimal working point, called flux-locked loop (FLL). Therefore, the output voltage of the FLL electronics is proportional to the magnetic flux change in the SQUID, and to the current through the feedback coil. Since the TES sensor is biased by constant steady current, ${I_{DB}}$, this operation mode is called ``dc-SQUID'' \cite{book:SQUID}. 

We can measure the current-voltage (I-V) characteristics of the NbSi TES sensors as the followings. At the superconducting state of the NbSi TES sensor, the resistance, ${R_{TES}=0 \Omega}$ at, for example, 30 mK; therefore, ${I_{bias}=0}$, ${I_{TES}=I_{DB}}$. In the mean while, we measure the sensitivity of the SQUID=320mV/${\mu}$A: for 1 ${\mu}$ A of detector bias current (${I_{DB}=I_{TES}}$) input, the voltage output (${V_{out}}$) reads 1.07V. Then we raise up the temperature to, for example, 135 mK, and we measure the voltage readout, ${V_{out}}$ by varying the detector bias current, ${I_{DB}}$, then we can calculate the current and the voltage of the NbSi TES sensor from the SQUID sensitivity and Eq. \ref{I_tes}:

\begin{equation}
I_{TES}=V_{out}/(320mV/\mu A),
\end{equation}

\begin{equation}
V_{TES}=I_{TES}\times R_{TES}=I_{TES} \times (\frac{R_{bias}\times I_{DB}}{I_{TES}}-R_{bias}).
\end{equation}

\begin{figure*}
\includegraphics{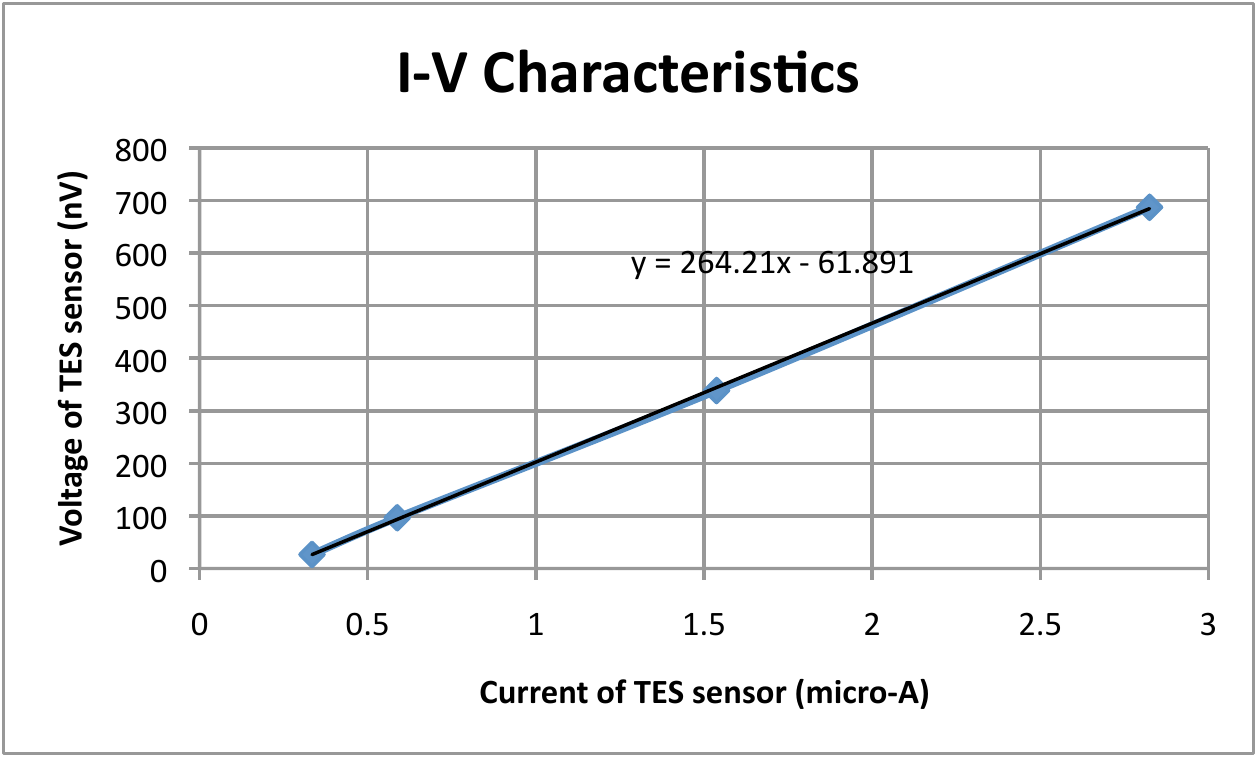}
\caption{\label{SQUID_IV} Current-voltage characteristics curve of the NbSi TES sensor with 50 nm thickness at 135 mK.}
\end{figure*}

From the slope of the I-V characteristics curve, we get the resistance of the NbSi TES sensor of 50 nm thickness, ${R_{TES}=264 m \Omega}$ at 135 mK, as shown in Fig. \ref{SQUID_IV}.

\begin{figure}
\includegraphics[width=6.5in,height=4in]{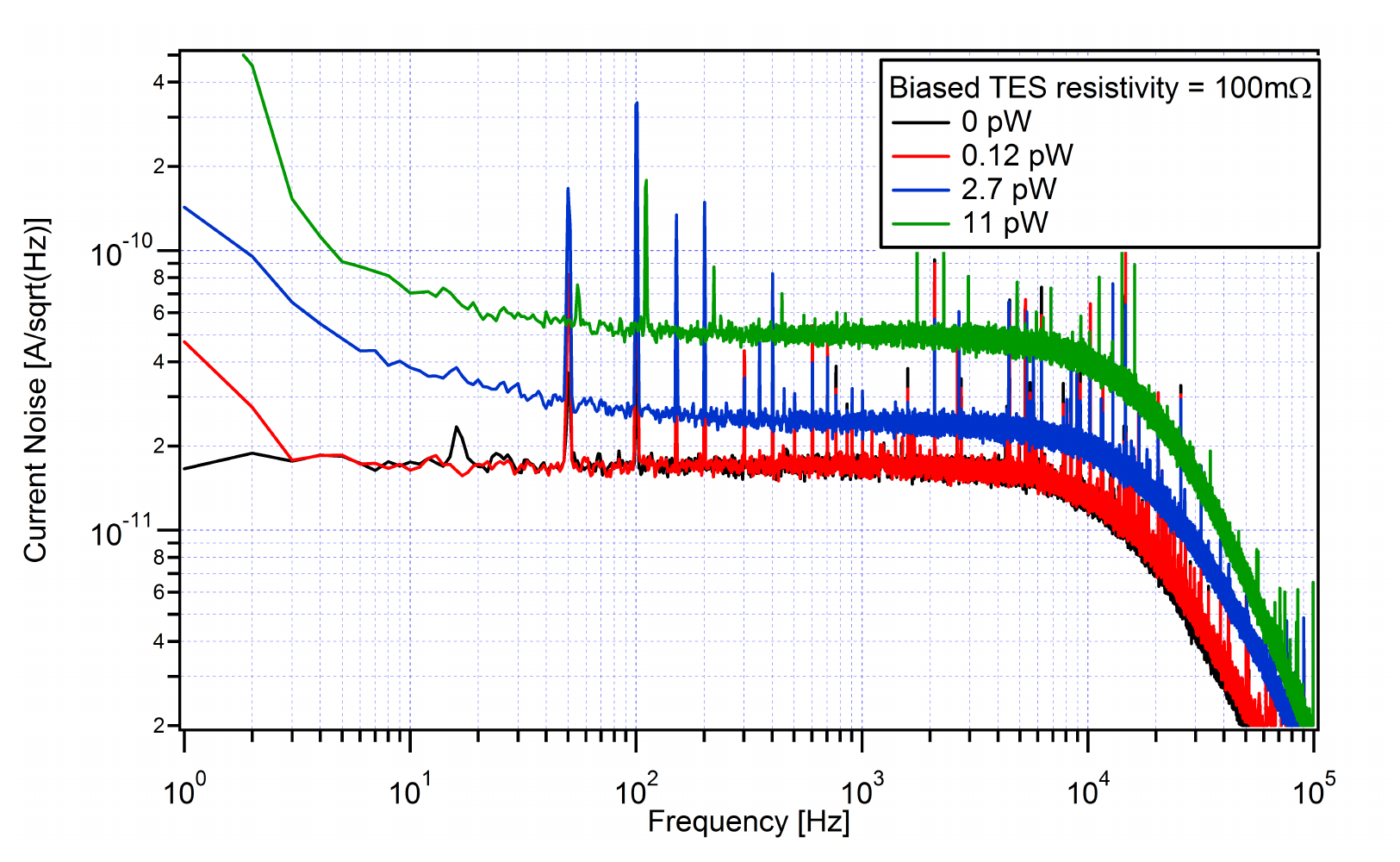}
\caption{\label{noise} Noise spectra of NbSi TES sensor.}
\end{figure}

The intrinsic noise sources of a bolometer include Johnson noise and phonon noise (i.e. thermal fluctuation noise), which can be described by the noise equivalent power (NEP). Johnson noise:
\begin{equation}
NEP_J^2=\frac{4kTR}{\Re ^2}[W^2/Hz]\propto T,
\end{equation}
where R is the electrical resistance at the temperature T, k is the Boltzman constant, ${\Re}$ is responsivity in V/W. The phonon noise:
\begin{equation}
NEP_{ph}^2=4kT^2G_d[W^2/Hz]\propto T^2,
\end{equation}
where ${G_d}$ is the conductance at uniform temperature T. The total intrinsic noise of the bolometer becomes:
\begin{equation}
NEP_{bol}^2=NEP_J^2+NEP_{ph}^2.
\end{equation}
Therefore, a TES should be operated at very low temperatures, below 100 mK, in order to reduce the sensor's phonon noise and Johnson noise to be smaller than the incident photon noise. Other noise sources are called ``excess noise''. Using the dc-SQUID, we measure the noise spectra of NbSi TES sensor, as shown in Fig. \ref{noise}. Johnson noise is about ${2\times 10^{-11}A/\sqrt{Hz}}$ at zero bias. At low bias we observe a flat Johnson noise while at higher bias the noise grows up proportionally to the current sensitivity, giving a constant signal to noise ratio. We observe a low frequency component in our noise spectra related to the temperature fluctuations of the cryostat and a high cut-off frequency at 20kHz coming from our low pass filter.

\section{\label{sec:level1}Conclusion}
We have developed sensitive and low noise bolometer matrices with NbSi TES sensors for the application of measuring the temperature fluctuations of the CMB (Cosmic Microwave Background radiation). The structure of bolometer matrices, transition edge sensors and fabrication process of our NbSi TES have been discussed. Our NbSi TES works not only as a temperature sensor but also as an absorber of radiation and due to its intrinsic thermal decoupling it does not need SiN membranes.

An electron-phonon coupling model is used to find out the electron-phonon coupling coefficients and the electron-phonon thermal conductance for different thickness in the NbSi thin film. The values are quite comparable with those found in metallic samples by other groups using different measurement methods, despite the fact that we are using the electron-phonon coupling model in the transition region of our TES samples. We also discuss about the performance of our NbSi TES bolometers. Their high sensitivity and low noise are very encouraging for Astroparticle detection experiments. 

\begin{acknowledgments}
We wish to acknowledge the support of the \dots.
\end{acknowledgments}

\newpage 

\bibliography{apssamp}

\end{document}